\documentclass[prd,preprint,superscriptaddress,a4paper,showpacs,showkeys,11pt,nofootinbib]{revtex4}
\usepackage{graphicx}
\usepackage{graphics}
\usepackage{dcolumn}
\usepackage{bm}
\usepackage{multirow}
\usepackage{tabularx}
\usepackage{hyperref}
\usepackage{commath}
\usepackage{epstopdf}
\usepackage[T1]{fontenc}
\usepackage[latin9]{inputenc}
\usepackage{geometry}
\usepackage{soul}
\usepackage{amsmath,amssymb,amsfonts}

\usepackage{color}

\def\be{\begin{equation}}
\def\ee{\end{equation}}

\providecommand{\ee}{e$^+$e$^-$}

\newcommand{\gaga}{$\gamma\gamma$}
\newcommand{\gaP}{$\gamma\mathbb P$}
\newcommand{\PP}{$\mathbb P\mathbb P$}

\makeatother

\newcommand{\pom}{\tt I\! P}

\begin{document}

%
%
\title{Challenging exclusive top quark pair production at low and high luminosity LHC}

\author{Daniel E. Martins}
\email[]{dan.ernani@gmail.com}
\affiliation{Instituto de F\'{\i}sica e Matem\'atica, Universidade Federal de
Pelotas (UFPel),\\
Caixa Postal 354, CEP 96010-090, Pelotas, RS, Brazil}

\author{Marek Tasevsky}
\email[]{Marek.Tasevsky@cern.ch}
\affiliation{Institute of Physics of the Czech Academy of Sciences, 
Na Slovance 2, 18221 Prague 8, Czech Republic}

\author{Victor P. Gon\c calves}
\email[]{barros@ufpel.edu.br}
\affiliation{Instituto de F\'{\i}sica e Matem\'atica, Universidade Federal de
Pelotas (UFPel),\\
Caixa Postal 354, CEP 96010-090, Pelotas, RS, Brazil}


\begin{abstract}
  The elastic production of top quark pairs in $pp$ collisions at low and high luminosity
  regimes is investigated in detail. We extend the study performed in Phys. Rev. {\bf D102}, 074014 (2020) ~\cite{ourfirst} which has
  demonstrated that the
  sum of two semi-exclusive $t\bar{t}$ production modes, namely in photon--Pomeron (\gaP) and
  Pomeron--Pomeron (\PP) interactions, can be experimentally measured when the $t\bar{t}$ system
  decays semi-leptonically, $t\bar{t}\rightarrow jjbl\nu_l\bar{b}$, both forward protons are tagged
  and a low amount of pile--up is present. In this study we focus on separating individual channels
  and a special attention is paid to the situation at high-luminosity LHC environment. We observe
  that the separation of the \PP\ and \gaga\ events is a challenging task, especially at high
  amounts of pile--up even with an optimistic 10~ps resolution of timing detectors. In contrast,
  the \gaP\ signal is relatively well separable from all backgrounds at low levels of pile--up,
  allowing us to discover the elastic $t\bar{t}$ production and probe, for the first time,
  the production of a top quark pair in the \gaP\ interactions.
  The diffractive photoproduction of such a complex system as the $t\bar{t}$ pair hence
  can be used not only to study diffractive properties of the scattering amplitude but also
  to search for new physics beyond the Standard Model, and consequently to be a solid part of the
  physics programme of forward proton detectors at LHC. 
\end{abstract}




\maketitle

\section{Introduction}\label{sec:intro}
\begin{figure}[t]
\begin{center}
\scalebox{0.5}{\includegraphics{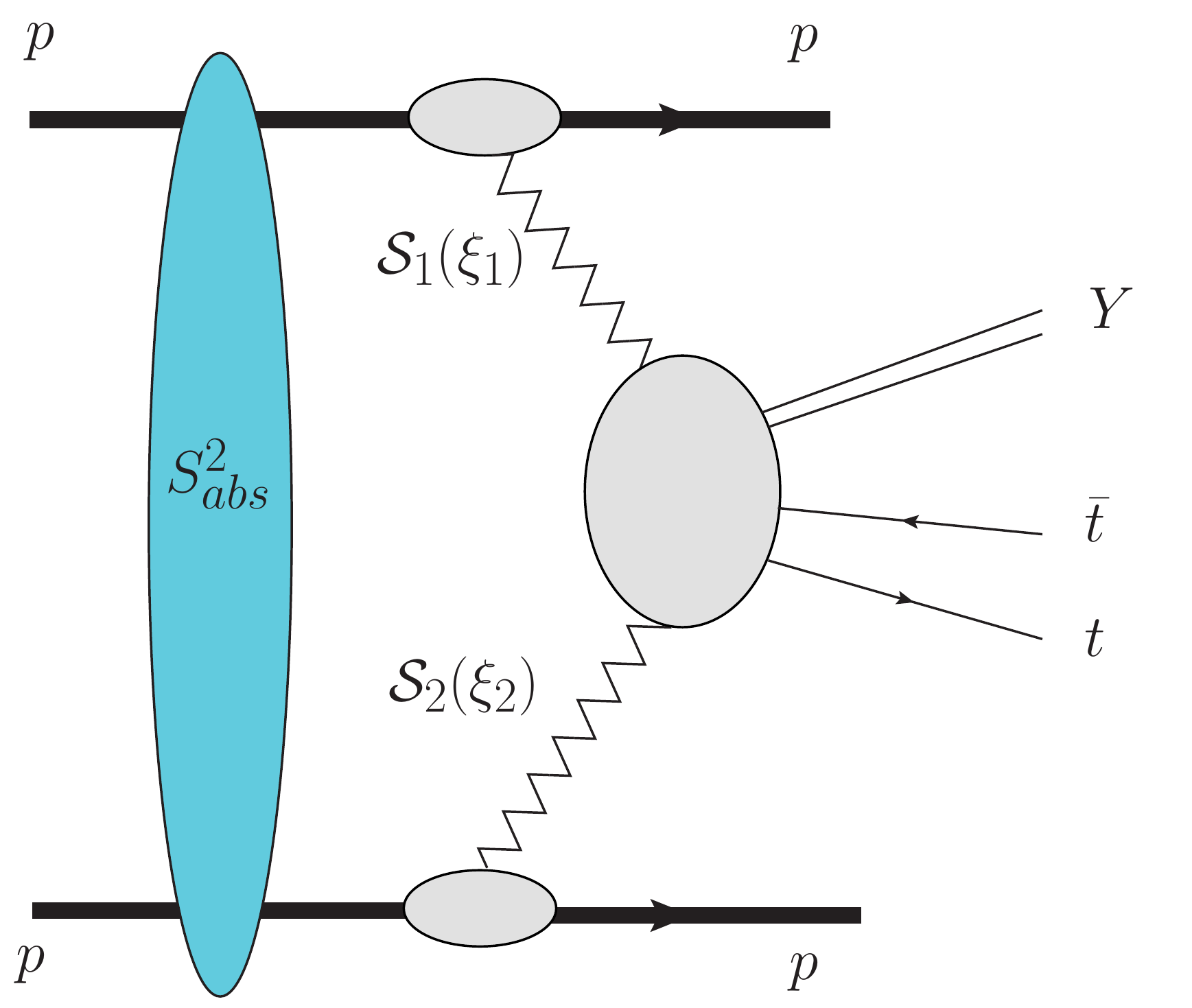}}
\caption{Elastic top pair production in $pp$ collisions, where ${\cal{S}}_1$ and ${\cal{S}}_2$ denote a
  color singlet object (photon ($\gamma$) or Pomeron ($\pom$)), $\xi_i$ is the fractional momentum loss
  of incident protons and $Y$ are the Pomeron remnants present in \gaP\ and \PP\ interactions.}
\label{Fig:diagram}
\end{center}
\end{figure}

The top quark plays a central role in the Standard Model (SM) and is generally considered to be an
excellent probe for new physics beyond Standard Model (BSM). It is the heaviest particle of the SM, with
a mass close to the scale of the electroweak symmetry breaking, implying that the top production and
decays at colliders are very sensitive to the presence of BSM phenomena (see e.g. Refs.~\cite{Husemann:2017eka,Fayazbakhsh:2015xba}).
Such aspects have strongly motivated studies of top quark production at the LHC, where top quarks are
produced with a high production rate in inelastic proton--proton collisions, where both incident protons
break up and a large number of particles is produced in addition to the top quarks (for recent
experimental results see, e.g. Refs.~\cite{Aad:2019mkw,Aad:2019ntk,Aad:2019hzw,Sirunyan:2018wem,Sirunyan:2017mzl,Sirunyan:2018goh}). 
It turned out that the analysis of top quark production in inelastic collisions generally involve serious
backgrounds, thus making the search for new physics a hard task. An alternative, recently proposed in
Ref.~\cite{ourfirst} (see also Ref.~\cite{Howarth:2020uaa}), is the study of top quark production in
$pp$ collisions characterized by intact protons in the final state. In what follows, this process will be
denoted as {\it elastic} top quark production and a schematic diagram is represented in
Fig.~\ref{Fig:diagram}. The basic idea is that the incident protons emit color singlet objects
${\cal{S}}_1$ and ${\cal{S}}_2$, which can be a quasireal photon $\gamma$ or a Pomeron $\pom$, the
protons remain intact and scatter with some energy loss $\xi$ in a very small angle from the beam pipe.
This way the top quark pair can thus be produced in \gaga, \gaP\ or \PP\ interactions. While the \gaga\
interaction is a purely exclusive process, where only the $t\bar{t}$ pair is present in the final state,
in \gaP\ and \PP\ interactions the pair is accompanied by Pomeron remnants (denoted by $Y$ in
Fig.~\ref{Fig:diagram}) when it is modeled as a quasi-real color singlet particle with partonic structure
and such processes are denoted as semi-exclusive. Elastic production is also characterized by the
presence of two rapidity gaps, i.e. two regions devoid of hadronic activity separating the intact very
forward protons from the central system. In principle, events associated to the elastic top quark
production can be clearly distinguished from the inelastic events by detecting the scattered protons in
spectrometers placed in the very forward region close to the beam pipe, such as the ATLAS Forward Proton
detector (AFP) \cite{Adamczyk:2015cjy,Tasevsky:2015xya} and the CMS--Totem Precision Proton Spectrometer
(CT--PPS) \cite{Albrow:2014lrm}, and selecting events with two rapidity gaps in the central detector.
However, in reality, the separation of elastic events is a challenge due to the presence of extra $pp$
interactions per bunch crossing, usually called pile--up, in high luminosity $pp$ collisions at the
LHC. The pile--up generates additional tracks that in general destroy the signature associated to two
rapidity gaps and increase the background stemming from the inelastic top quark pair produced in a
different primary vertex.

In our previous article~\cite{ourfirst}, we have performed the first comprehensive study of the elastic
top quark pair production in $pp$ collisions at $\sqrt{s} = 13$ TeV taking into account the current
detector acceptances and resolutions as well as the pile--up effects expected for the next run of the
LHC. Good prospects for observing the elastic signal over a mixture of inclusive and combinatorial
background were achieved for all luminosity scenarios considered, although a good separation between the
two is observed for rather low amounts of pile--up, typically lower than 50.
In particular in Ref.~\cite{ourfirst} we showed that the separation is feasible for a sum of two
semi-exclusive production modes, \gaP\ and \PP, when the $t\bar{t}$ system decays semi-leptonically,
$t\bar{t}\rightarrow jjbl\nu_l\bar{b}$, both forward protons are tagged and a low amount of pile--up is
present. Our goal in this paper is to improve and extend the analysis performed in Ref.~\cite{ourfirst}
by considering additional exclusivity cuts and utilizing the time-of-flight (TOF) detectors for
suppressing the combinatorial background coming from pile-up. We will focus on  separating the
$t\bar{t}$ final state produced in semi-exclusive (\gaP\ and \PP) or even exclusive (\gaga) processes
from backgrounds at the low -- and high -- luminosity LHC environments. 
As we will demonstrate below, the separation of the \PP\ and \gaga\ events will be a challenging task,
especially at high amounts of pile--up even with an optimistic 10~ps resolution of timing detectors. In
contrast, the events associated to the top quark production in \gaP\ interactions can be relatively well
separable from all backgrounds at low levels of pile--up. Such results indicate that the elastic
$t\bar{t}$ production can be discovered in a near future at the LHC and that an experimental analysis of
this process will be able to probe, for the first time, the diffractive photoproduction of top quark pairs. Moreover,
such a perspective opens the possibility to search for new physics beyond the Standard Model in this
process.

This paper is organized as follows. In next Section, we present a brief review of the formalism for the
elastic top pair production $pp$ collisions. In Section~\ref{sec:exp} we discuss details of the selection
of events and cuts implemented in our analysis, concentrating on collisions at $\sqrt s = 13$~TeV in the
low -- and high -- luminosity regimes. In Section~\ref{sec:results} we present our results for the
distributions of the fractions of proton momentum loss and invariant mass and our predictions for the
effective cross sections. Moreover, the significance for the distinct processes is estimated considering
different amounts of pile--up. Finally, in Section~\ref{sec:sum} we summarize our main findings.

\section{Formalism}\label{sec:form}
For completeness, in this Section we will present a brief review of the formalism needed to describe the
elastic top pair production in $\gamma \gamma$, $\gamma \pom$ and $\pom \pom$ interactions and refer the
reader to Ref. \cite{ourfirst} for a more detailed discussion. 
A well known aspect is that an ultra relativistic proton acts as a source of almost real photons and that
the associated photon spectrum can be computed using the equivalent photon approximation \cite{epa}. In a
similar way, one can associate a Pomeron flux to the incident protons, with the Pomeron being usually
modeled as a quasi-real color singlet particle with a partonic structure \cite{IS}.
As a consequence, the total cross section for the elastic top pair production can be factorized in 
terms of the equivalent flux of photons and Pomerons into the proton projectiles and the photon--photon,
photon--Pomeron or Pomeron--Pomeron production cross section. Denoting by ${\cal{S}}_{1,2}$ the generic
color singlet object, the cross section can be represented in a schematic way as in
Fig.~\ref{Fig:diagram} and expressed as follows
\begin{eqnarray}
\sigma(pp \rightarrow p \otimes t\bar{t} \, Y \otimes p) \propto S^2_{abs} \times f_{{\cal{S}}_1}(\xi_1) \times f_{{\cal{S}}_2}(\xi_2)  \times \hat{\sigma}({\cal{S}}_1 {\cal{S}}_2 \rightarrow t \bar{t}) \,\,,
\label{gene}
\end{eqnarray}
where $\otimes$ represents the presence of a rapidity gap in the final state, $\xi_i$ is the fraction of
the incident proton energy carried by the color singlet object ${\cal{S}}_i$ and $f_{{\cal{S}}_i}$ is the
equivalent photon or Pomeron distribution of the proton. Moreover, $Y$ denote the Pomeron remnants
present in \gaP\ and \PP\ interactions and $S^2_{abs}$ is the rapidity gap survival factor, which takes
into account additional soft interactions between the incident protons which leads to an extra production
of particles that destroy the rapidity gaps in the final state \cite{bjorken}.

In order to estimate the total cross sections and associated distributions for the elastic top pair
production, one has to assume a model for the equivalent photon and Pomeron distributions as well as for
the survival factor. Following Ref.~\cite{ourfirst}, we will use the photon flux from Ref.~\cite{epa},
where an analytical expression is derived. Moreover, the Pomeron distribution is expressed in terms of
the diffractive parton distributions, whose evolution is described by the DGLAP evolution equations and
are determined from events with a rapidity gap or intact proton, mainly at HERA collider~\cite{pomeron}.
As in Ref.~\cite{ourfirst} we will express these quantities assuming the validity of the resolved Pomeron
model~\cite{IS} and will describe the diffractive parton distributions by the parameterization obtained
by the H1 Collaboration at HERA, denoted as the fit A in Ref.~\cite{H1diff}.
As emphasized in Ref.~\cite{ourfirst}, the treatment of $S^2_{abs}$ for \gaga, \gaP\ and \PP\ interactions
is still a theme of intense debate due to the non-perturbative nature of the additional interactions
(see, e.g. Ref.~\cite{review_martin}). 
We will assume  that the hard process associated to the top pair production occurs on a short enough
time scale such that physics that generate the additional particles can be factorized, which allows us
to parametrize the absorption effects in terms of a global constant factor. While for the photon--photon
and photon--pomeron collisions the contribution of the additional soft interactions is expected to be
small due to the long range of the electromagnetic interaction, it is non-negligible for pomeron--pomeron
collisions and imply the violation of the QCD hard scattering factorization theorem for diffraction in
$pp$ collisions \cite{collins}. 
In our analysis, following Ref.~\cite{ourfirst}, we will assume $S^2_{abs} = 1$ for \gaga\ and \gaP\
interactions. Recent studies \cite{Dyndal:2014yea,Harland-Lang:2020veo,Harland-Lang:2021ysd} have derived
a smaller value for
the survival factor in \gaga\ interactions. As a consequence, our predictions can be considered to be
an upper bound for the elastic top pair production in photon--photon interactions. In contrast, for
\gaP\ interactions, such assumption is a very good approximation, since the resolved Pomeron model is
able to describe the diffractive charm photoproduction measured at HERA. On the other hand, the modeling
and magnitude of $S^2_{abs}$ for \PP\ interactions are still open questions, with its value being typically
of the order of 1 -- 5\% for LHC energies.
As in previous studies for single and double diffractive production \cite{potterat,antoni,antoni2,dimuons,palota} we will assume  $S^2_{abs} = 0.03$ for pomeron--pomeron interactions as predicted in
Ref.~\cite{KMR} but we will also try to address the question of what is the impact on our results if
a larger value is considered. 


\section{Experimental procedure}\label{sec:exp}
The procedure used to separate the elastic signal from backgrounds, represented mainly
by pile-up, is based on that described in detail in Ref.~\cite{ourfirst}, so here we only remind the
main steps and then we will describe two new cuts introduced in this analysis. In Ref.~\cite{ourfirst}
we have demonstrated that
background processes $\gamma\gamma\rightarrow WW$ and $\gamma\mathbb P\rightarrow Wt$ can safely be
neglected in the case of our signal induced by \gaP\ and \PP\ interactions.
That allows us to consider as backgrounds for each signal process the inclusive production of
$t\bar{t}$ pair and the other two elastic processes, all overlaid with pile-up
(as an example, in the case of $\gamma\mathbb P\rightarrow t\bar{t}$ signal, the backgrounds are
inclusive $t\bar{t}$, $\gamma\gamma\rightarrow t\bar{t}$ and $\mathbb{PP}\rightarrow t\bar{t}$,
all evaluated including pile-up effects).
To estimate the statistical significance, $\sigma$, and signal to background ratio, S/B, we consider
three luminosity scenarios in terms of $\langle \mu \rangle$ and $\cal L$ where
$\langle \mu \rangle$ represents the average number of pile-up interactions per event (or the
instantaneous luminosity) and $\cal L$ is the integrated luminosity. We assume $\cal L$ to be 10,
300 and 4000~fb$^{-1}$ for $\langle \mu \rangle =$~5, 50 and 200, respectively, the last one
corresponding to the assumed conditions at HL-LHC. It is appropriate to note that compared to the
approach adopted in Ref.~\cite{ourfirst}, namely to use a simple formula $S/\sqrt{B}$ to estimate
significances, in this study we use the formula based on Asimov data set~\cite{Cowan:2010js} which
reflects the reality more reliably and which reduces to the simple ratio above if $S \ll B$.

The separation of the elastic $t\bar{t}$ signal from backgrounds at proton-proton collisions
at centre-of-mass system energy of $\sqrt s = 13$~TeV proceeds in three steps: first we select the
central system as
in the inclusive processes, then we apply exclusivity criteria and finally we add two new cuts,
with the aim to separate individual processes from each other. In the first step,
we concentrate on the so-called semi-leptonic $t\bar{t}$ decay,
$t\bar{t}\rightarrow jjbl\nu_l\bar{b}$, in the second step, we require both forward protons to be
detected by Forward Proton Detectors (FPDs), to be processed by Time-of-Flight (ToF) detectors with
an assumed resolution of 10~ps and the
decay products of the $t\bar{t}$ pair to be accompanied by a low number of particles. The third step
consists of cuts on the proton transverse momentum and a 2-dimensional cut in the
$(m_X, m_{t\bar{t}})$ plane where $m_X$ is the missing mass evaluated from forward proton measurements
and $m_{t\bar{t}}$ is the mass of the $t\bar{t}$ system measured by the central detector. 

All three signal processes are generated using the Forward Physics Monte Carlo (FPMC) \cite{fpmc}, while the
inclusive $t\bar{t}$ background is generated by MadGraph~5 \cite{madgraph}. Each event is then properly
mixed with such a number of pile--up interactions that corresponds to the studied
luminosity scenario (e.g. for the ($\langle \mu \rangle ,\cal L$) = (5,10) point, the actual
number of overlaid pile--up events is coming from a Poisson distribution with the mean of 5.0). 
A sample for the pile--up mixing consists of 200 thousands of minimum bias events generated by
PYTHIA~8 \cite{Sjostrand:2014zea}  including multi-parton interactions. Detector effects are incorporated and this pile--up
mixing is done using Delphes~3.5~\cite{delphes3} with input cards with CMS detector specifications.
Where available, ATLAS cards are used for systematic studies. For $\langle \mu \rangle = 5$ and 50,
Delphes package provides cards with both, the ATLAS and CMS parameters, while for
$\langle \mu \rangle = 200$, we used a CMS card tuned in recent HL-LHC studies~\cite{Michele_card}. 
For all, the elastic and inclusive $t\bar{t}$
processes, the mass of the top quark is set to the value of
174.0~GeV. For FPDs we assume a fully efficient reconstruction in the range
$0.015 < \xi_{1,2} < 0.15$, where $\xi_{1,2} = 1 - p_{z 1,2}/E_{\rm beam}$ is the
fractional proton momentum loss on either side of the interaction point
(side 1 or 2) and $p_{z 1}$ is the longitudinal momentum of the scattered proton on the side 1.
This, in principle, allows one to measure masses of the central
system by the missing mass method, $m_X\!=\!\sqrt{\xi_1\xi_2s}$, starting from
about 200~GeV. 
Large event samples of the aforementioned processes have been generated, each of at least 200,000
events, corresponding ---in the case of signal processes--- to integrated luminosities that sufficiently
exceed the assumed ones in the three considered luminosity scenarios. 

All cuts used in the first two steps are elaborated in Ref.~\cite{ourfirst}. One can find there 
details about jet definition, b-tagging, about lepton definition and isolation, about
number of tracks originating in a narrow region around the primary vertex and far from all
objects in the final state (so called z-vertex veto). Here we would like to only remind how we
make use of the ToF detector.
First we evaluate a probability to see an intact proton from one minimum bias event in the FPD
$\xi$-acceptance on one side and after, eventually, applying the proton $p_T$ cut, $P_{\rm ST}$.
Based on this
probability, we then calculate combinatorial factors representing rates of fake double-tagged
events for the three studied working points of $\langle \mu \rangle$ and for various options for
proton $p_T$ cut. Finally the ToF suppression factors are enumerated for each combination of
$\langle \mu \rangle$ and proton $p_T$ cut, under the assumption that the time resolution of ToF is
$\sigma_{\rm ToF} =$~10~ps and the signal is collected in a $\Delta t = \pm 2\sigma_{\rm ToF}$ window. More details
about the
ToF method and the $\langle \mu \rangle$-dependence of this combinatorial background can be found
in Refs.~\cite{Tasevsky:2014cpa,Harland-Lang:2018hmi,ToFperformance}. 
Given the large $\Delta t$ collection window, the efficiency of the signal collection at
$\langle \mu \rangle = 0$ is close to 100\%. In this study we assume a very good granularity of the
ToF detector and no $\langle \mu \rangle$ dependence of the signal or background collection efficiency.

\begin{figure}[htbp!]
\includegraphics[width=0.329\textwidth,height=4.7cm]{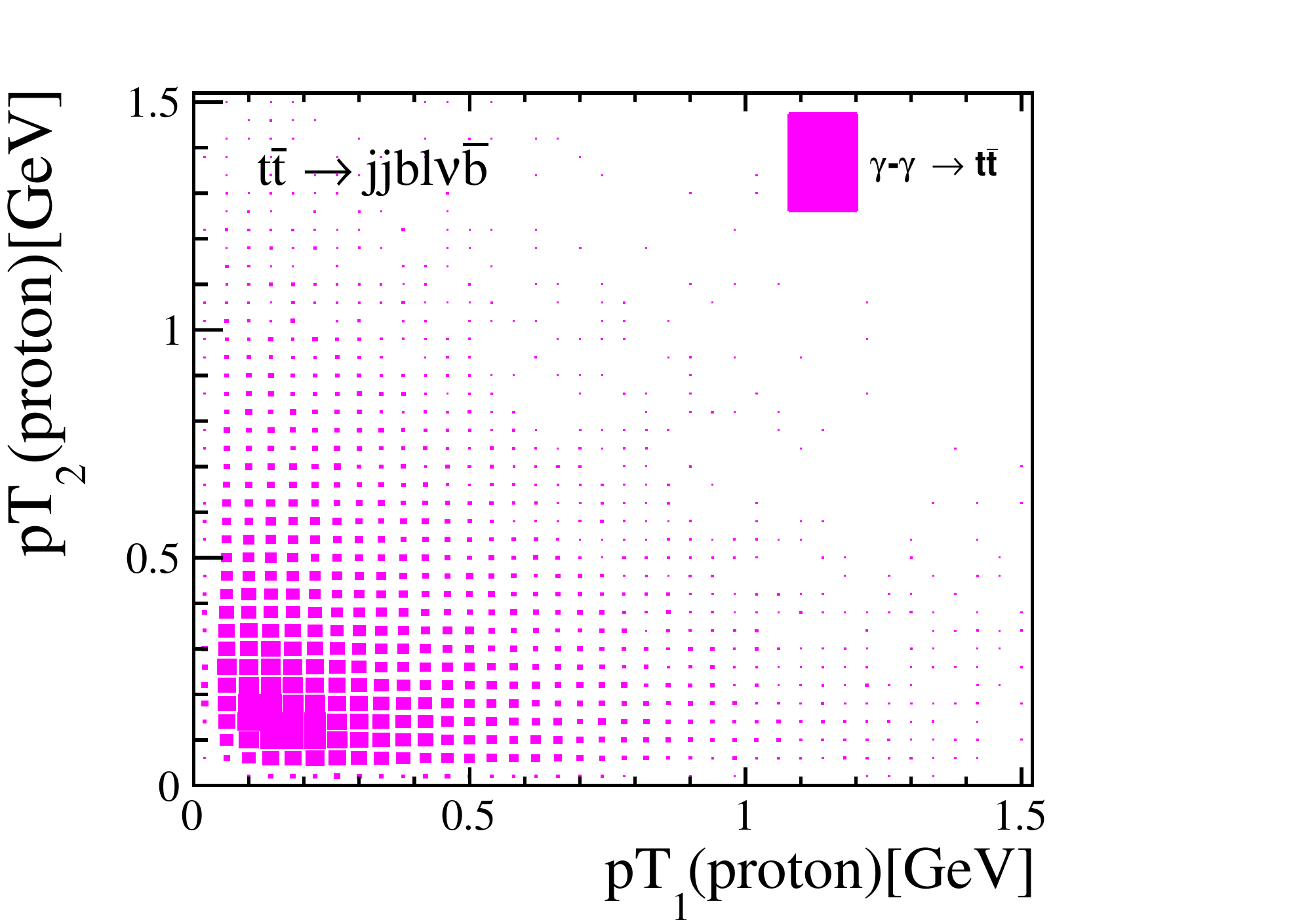}
\includegraphics[width=0.329\textwidth,height=4.7cm]{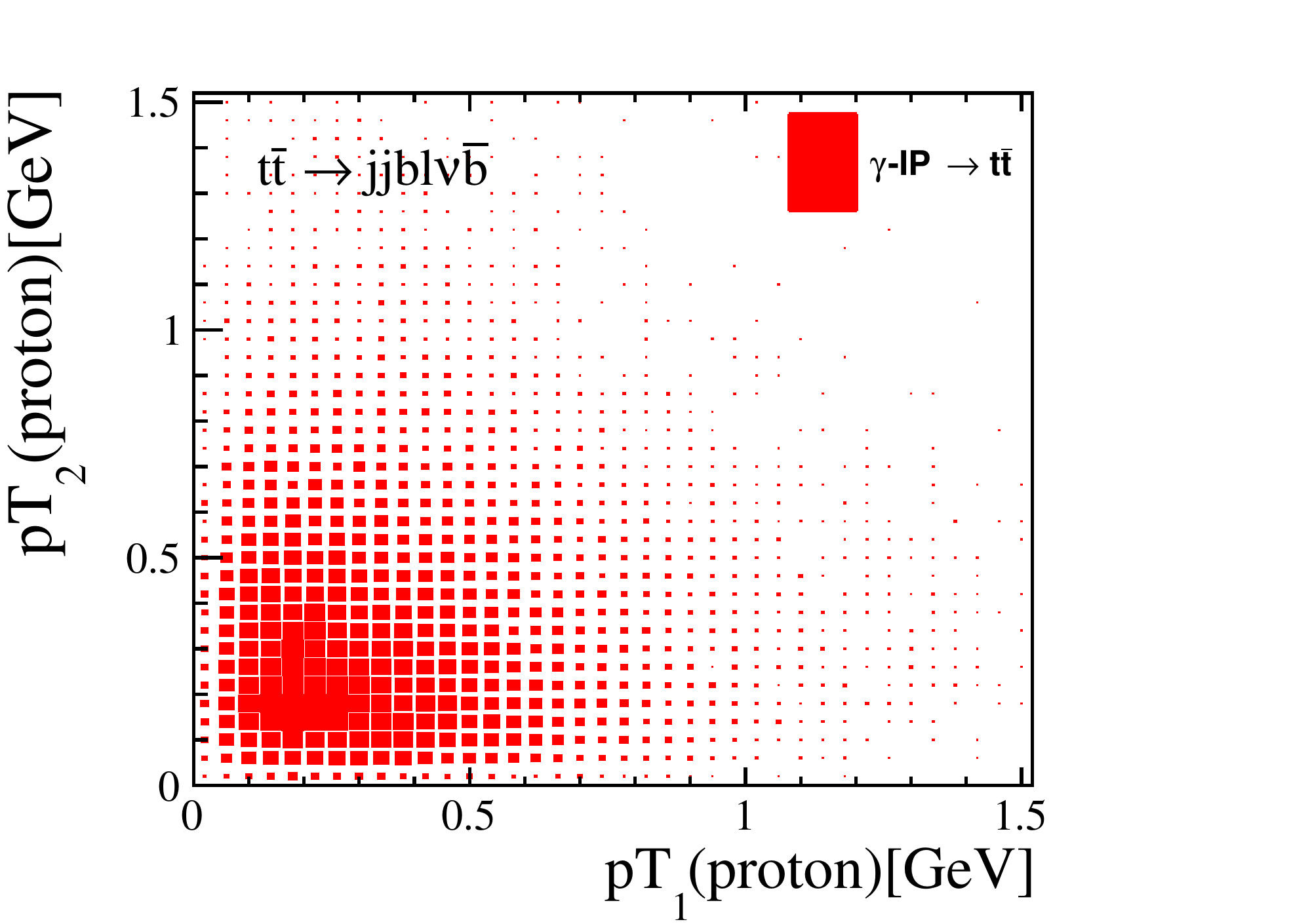}
\includegraphics[width=0.329\textwidth,height=4.7cm]{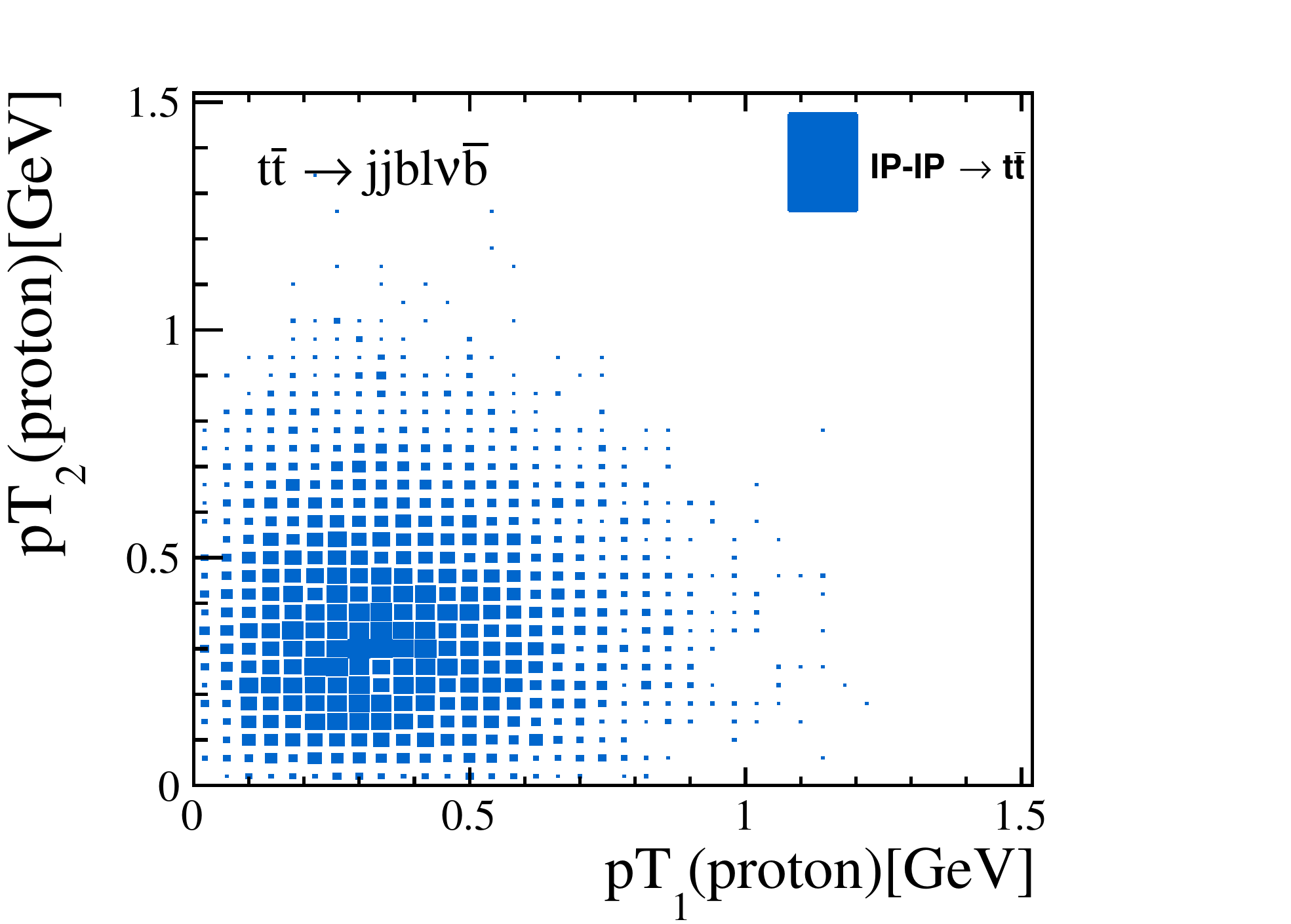}
\caption{2-dimensional distribution of the transverse momentum of proton on one side from the
  interaction point versus that on the opposite side, after applying cuts in Table~\ref{tab:cuts}
  up to the row corresponding to the $\xi$ acceptance inclusively and without considering pile-up
  effects. Predictions for the signal processes
  (\gaga\ on the left, \gaP\ in the middle and \PP\ on the right) are obtained with FPMC and are
  scaled by effective cross sections corresponding to the set of applied cuts.}
\label{proton_pt}
\end{figure}
\begin{figure}[htbp!]
\includegraphics[width=0.45\textwidth,height=6cm]{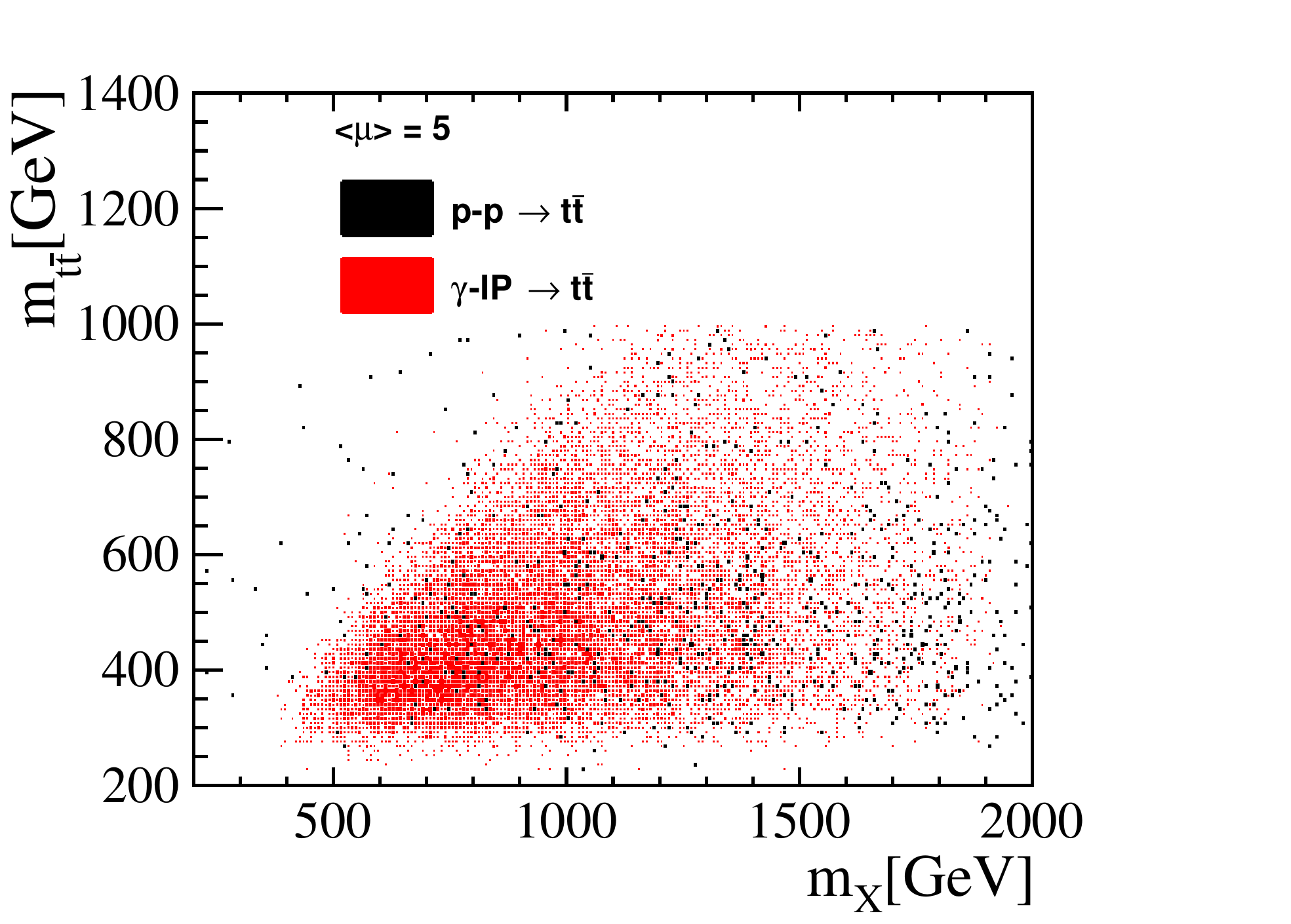}
\includegraphics[width=0.45\textwidth,height=6cm]{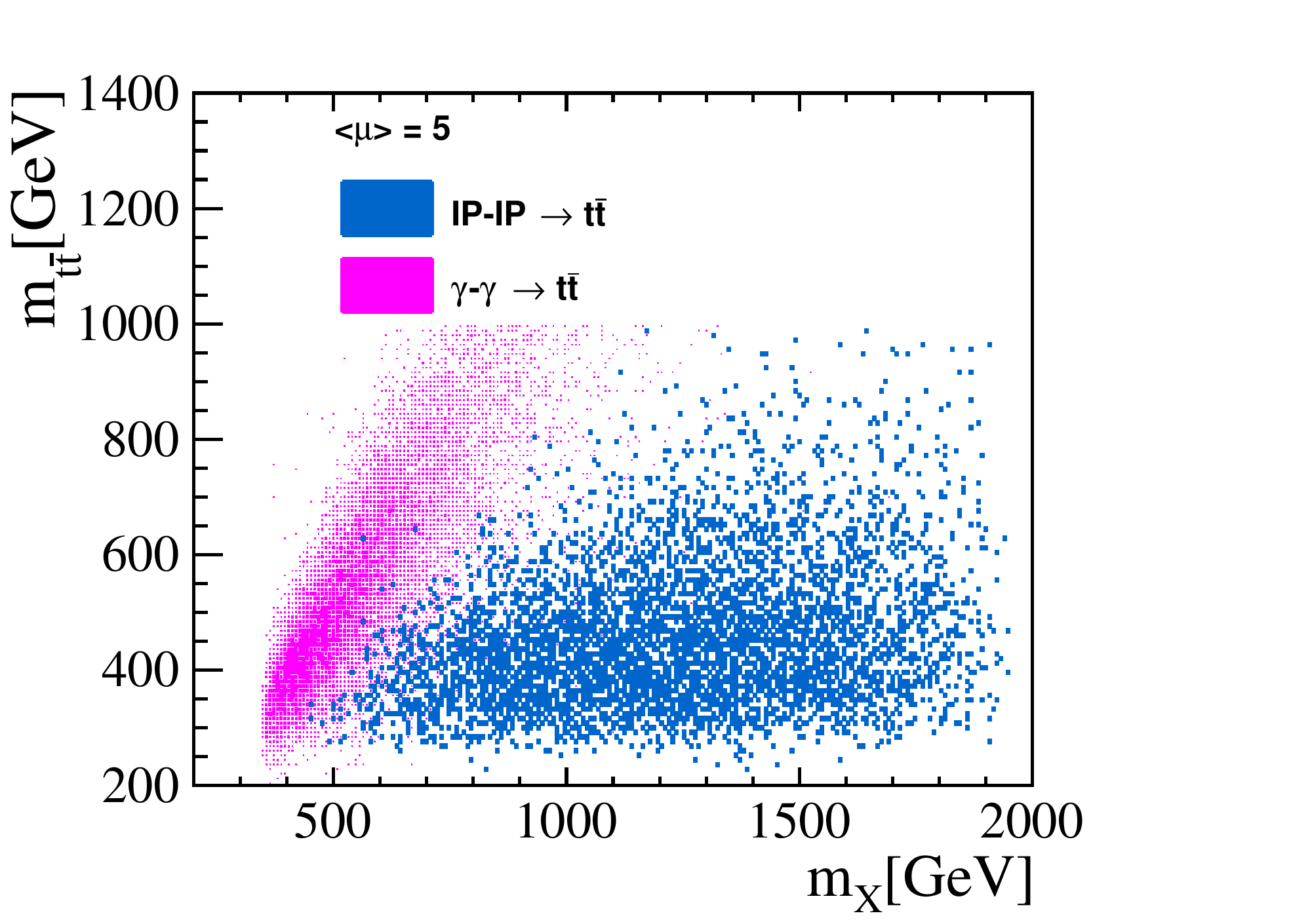}
\caption{2-dimensional distribution of the central $t\bar{t}$ system mass measured by forward proton
  detectors ($m_X$) versus that measured by central detector ($m_{t\bar{t}}$) after applying all cuts
  in Table~\ref{tab:cuts} up to the row corresponding to the $\xi$ acceptance inclusively.
  Predictions, scaled by effective cross sections
  corresponding to the set of applied cuts, are obtained with FPMC for processes \gaga, \gaP\ and
  \PP, while MadGraph~5+PYTHIA~8 is used for the inclusive $t\bar{t}$ production. All processes are
  mixed with pile-up with $\langle \mu \rangle = 5$.}
\label{2Dcut}
\end{figure}

In the third step, we apply two additional cuts with the aim to separate individual processes
from each other. The cut on proton $p_T$ is potentially a powerful cut if pile--up is not present
since it helps to separate
the elastic processes among each other on one hand and all elastic
processes from the background formed by the inclusive production plus pile-up on the other hand.
The reason for the former is a different $p_T$ spectrum of proton which emits photon from
that which emits Pomeron, see Fig.~\ref{proton_pt}. The reason for the latter is a decrease of the
rate of fake double-tagged events as a natural consequence of a decrease of the probability
$P_{\rm ST}$ whenever additional cuts on the forward proton are imposed.
Another potential background rejector is based on the fact that the individual processes studied
here occupy different regions in a 2-dimensional $(m_X, m_{t\bar{t}})$ plane where $m_X$ is the
missing mass evaluated from forward proton measurements and $m_{t\bar{t}}$ is the mass of the
$t\bar{t}$ system measured in the central detector. The situation at $\langle \mu \rangle = 5$ is
illustrated in Fig.~\ref{2Dcut} which suggests a relatively good potential for separation of
the \gaga\ signal from all other processes in general. When plotting these distributions, the
effective cross sections of the mix of inclusive and pile-up events are already scaled by the
corresponding rates of fake double tagged events and by ToF suppression factors.

All the cuts considered in our analysis are summarized in Table~\ref{tab:cuts}. Basically, we require
the following:
\begin{itemize}
\item In total at least four not-overlapping jets with $E_{\rm T,jet} > 25$~GeV
  and $|\eta_{\rm jet}| < 2.5$ (for HL-LHC, we consider the extended coverage $|\eta| < 4.0$ of the
  upgraded tracker).
\item At least one electron or muon ($\tau$ decays included)
  with $E_{\rm T,l} > 25$~GeV and $|\eta_{\rm l}| < 2.5\ (4.0)$ isolated from all four
  jets, $\Delta R_{\rm l,j} > 0.2$.
\item At least two b-tagged jets. A jet is b-tagged if a B-hadron (generator
   level) or a b-quark (detector level) is found inside the jet.
\item Exactly two forward protons, each in the FPD acceptance $0.015 < \xi_{1,2} < 0.15$.
\item Number of tracks with $p_{\rm T,trk} > 0.2$~GeV and $|\eta_{\rm trk}| < 2.5\ (4.0)$
  in the distance {$|z_{\rm trk}-z_{\rm vtx}|<1$~mm from the primary vertex and}
  $\Delta R_{\rm trk,j} > 0.4$ from the four jets and
  $\Delta R_{\rm trk,l} > 0.2$ from one lepton must be smaller than a given value X.
\item $p_T$ of each proton detected in FPD must be smaller than a given value Y.
\item Optimal area in the ($m_X, m_{t\bar{t}}$) plane is found for each signal process.
  We proceed in a rather simple manner and select --- merely on a visual basis --- the optimal
  area as a continuous part of the 2D plane with an advantageous S/B ratio. 
  For each process we tried several areas and kept the one giving the best significance but
  it is clear that better results can be obtained if more sophisticated methods are used.
  One can usually try to find Fischer discriminant or to perform regression analysis or,
  if more variables can be used to separate signal from background, one can make use of machine
  learning techniques, e.g. Boosted Decision Trees and Neural Networks.
  \end{itemize}

\begin{table}[t]
\begin{tabular}{||c||} 
\hline
Cut \\ \hline\hline
$N_{\rm jet} \ge 4~(E_T>25~{\rm GeV}, |\eta|<2.5\ (4.0))$ \\ \hline
$N_{e/\mu} \ge 1~(E_T>25~{\rm GeV}, |\eta|<2.5\ (4.0))$ \\ \hline
$\Delta R (\rm{e/\mu,jet})> 0.2$ \\ \hline
$N_{\rm b-jet} \ge 2$ \\ \hline
$0.015<\xi_{1,2}<0.15 $ \\ \hline
$N_{\rm trk} (p_{\rm T} > 0.2~{\rm GeV}, |\eta| < 2.5\ (4.0), {|\Delta z|<1~{\rm mm}}) \leq X $\\ \hline
$p_T^{\rm proton} < Y$ \\ \hline
Optimal area in the ($m_X, m_{t\bar{t}}$) plane \\ \hline 
\end{tabular}
\caption{Cuts used in this analysis. The extended $\eta$ coverage is considered only for the HL-LHC
scenario.}
\label{tab:cuts}
\end{table}


\section{Results}\label{sec:results}
Areas of population for individual processes shown in Fig.~\ref{2Dcut} clearly demonstrate that the process
\gaga\ can in principle be isolated well from all other processes.
Nevertheless due to its extremely low effective cross section, in order to collect a reasonable event yields,
we have to only consider large integrated and hence instantaneous luminosities and thus consider only
the luminosity scenarios ($\langle \mu \rangle ,\cal L$) = (50, 300) and (200, 4000). Then the good isolation seen at low values of
$\langle \mu \rangle$ is completely washed out
due to the inclusive background points scattered all around the area, as expected from flat mass
distributions for this type of background when pile-up is not negligible (see Figs.~2 and 3 in
Ref.~\cite{ourfirst} for masses larger than 500~GeV). The other additional cut on proton $p_T$ turns
out to be futile in improving statistical significance. The largest observed values are around 0.1 for
no cut on proton $P_T$ and $N_{\rm trk} \le 15$ for both $\langle \mu \rangle =50$ and
$\langle \mu \rangle =200$.

As far as the \gaP\ process is concerned, the extraction of an optimal area in the
$(m_X, m_{t\bar{t}})$ plane turns out to be useful only marginally when $\langle \mu \rangle =5$ and
it was abandoned completely at higher amounts of pile-up for the same reasons as mentioned above.
In general, the forward proton $p_T$ cut does not improve the situation since --- unlike for the
\gaga\ and \PP\ processes --- the $p_T$ spectrum is broader due to the fact that one proton emits
photon and the other Pomeron. At $\langle \mu \rangle =5$, this
signal is well observable over all backgrounds, with a significance of 3.4 for
$N_{\rm trk} \le 25$. The significances are lower for larger amounts of pile-up. At
$\langle \mu \rangle = 50$ a maximum significance of 2.1 is obtained for $N_{\rm trk} \le 25$ and at
$\langle \mu \rangle = 200$, a maximum significance of 2.3 is achieved for $N_{\rm trk} \le 30$
when the extended $|\eta| < 4.0$ coverage for the upgraded tracker is taken into account.
For the last case, we studied three scenarios regarding the z-veto region and minimum $p_T$ of
tracks counted inside it, namely 1~mm and 0.2~GeV (the nominal configuration used for lower
$\langle \mu \rangle$ values), 0.5~mm and 0.5~GeV, and 0.2~mm and 0.5~GeV, and we observed
practically no improvement. Very similar numbers are obtained with the ATLAS card where available
($\langle \mu \rangle = 5$ and 50). 

\begin{figure}[t]
\includegraphics[width=0.45\textwidth,height=6cm]{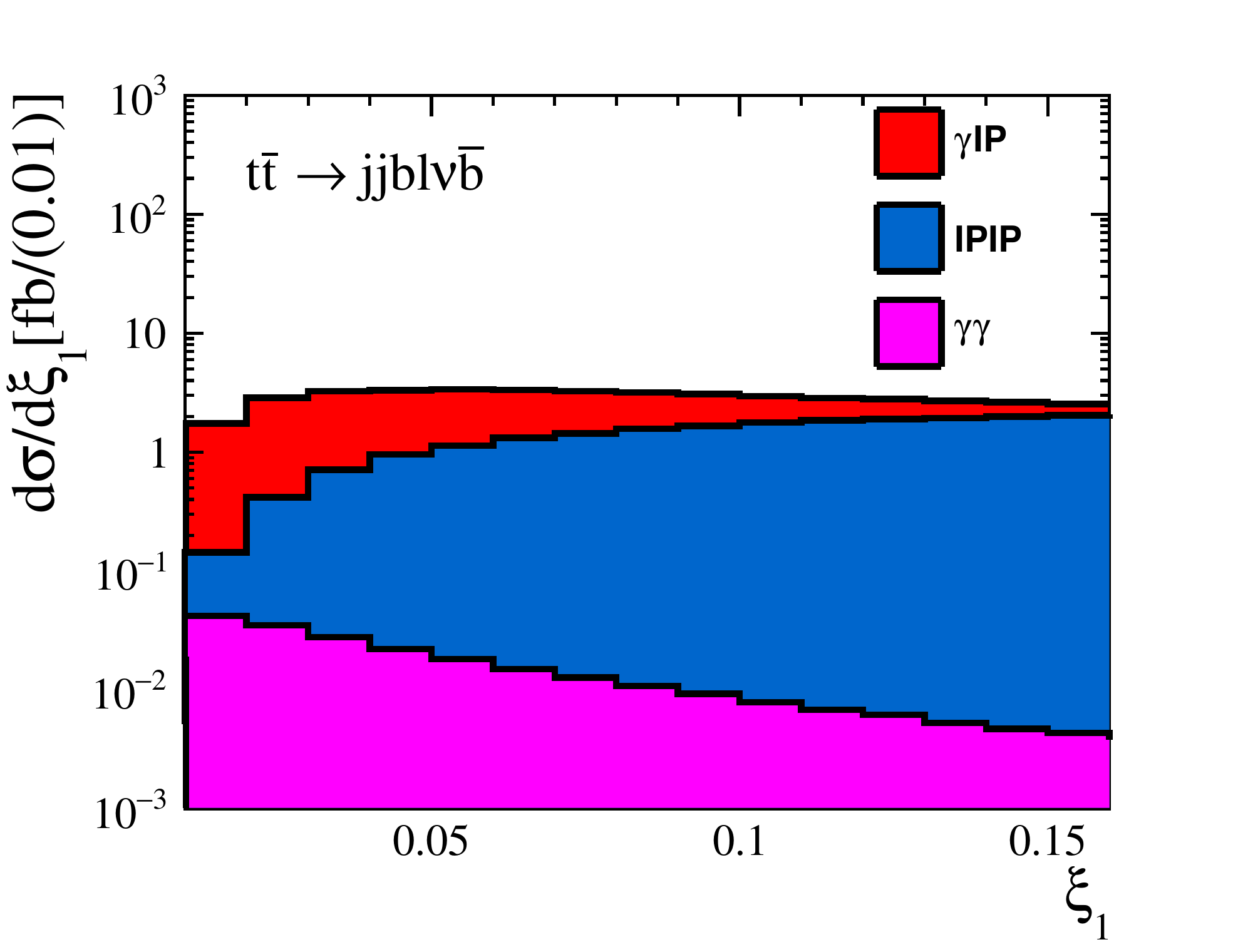}
\includegraphics[width=0.45\textwidth,height=6cm]{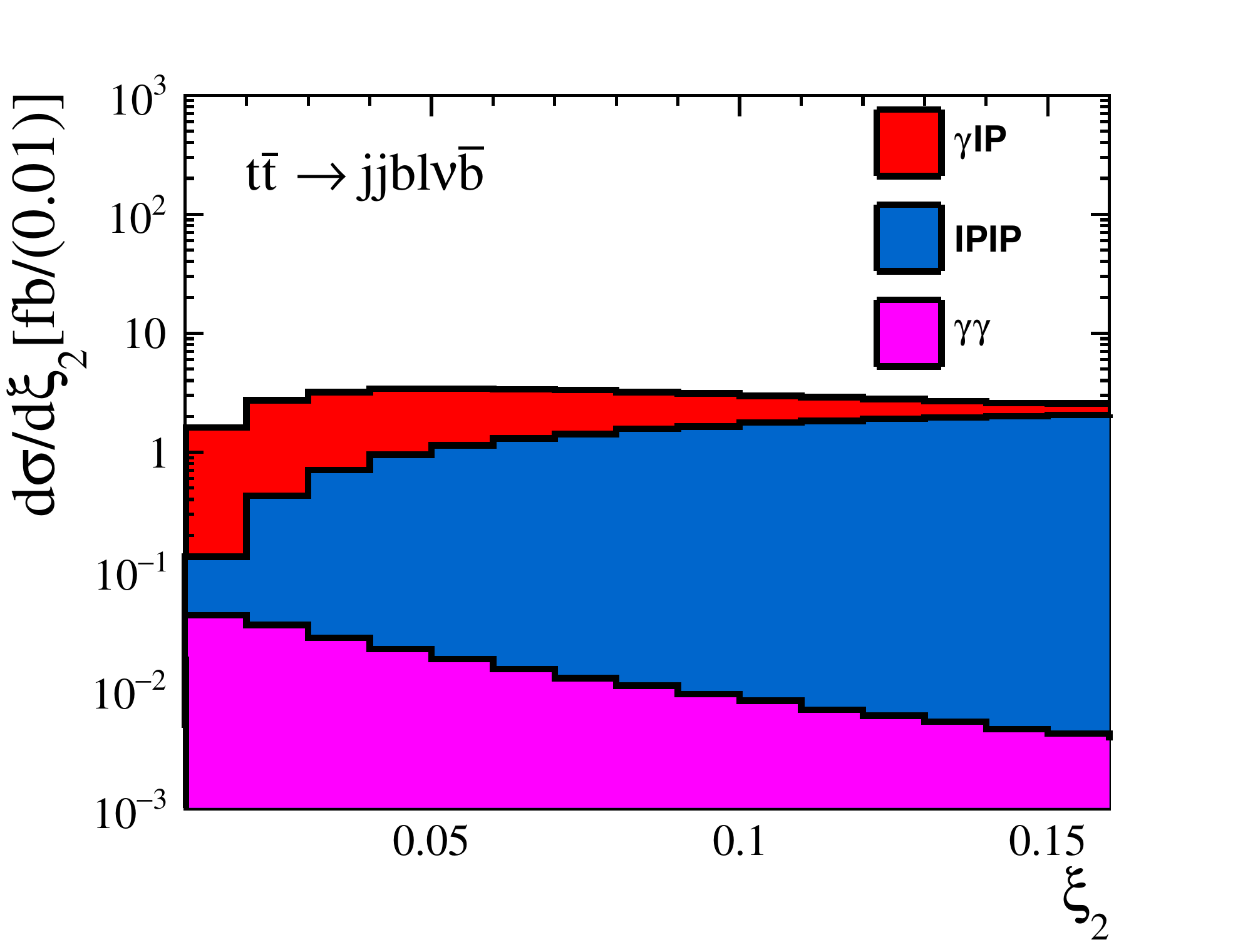}
\caption{Distributions of fractions of proton momentum loss on each side, after applying cuts in
  Table~\ref{tab:cuts} up to the row corresponding to the number of b-tagged jets inclusively and
  without considering pile-up effects. Predictions for the signal processes \gaga, \gaP\ and \PP\
  are obtained with FPMC and are scaled by effective cross sections corresponding to the set of
  applied cuts.}
\label{xidist}
\end{figure}

Only modest significances are obtained when trying to separate the \PP\ process. Although here
the proton $p_T$ cut plays a more important role than searching for the optimal area in the
$(m_X, m_{t\bar{t}})$ plane, its effect is still rather marginal and the best significance of 0.9
was achieved for no cut on proton $p_T$ and $N_{\rm trk} \le 20$. At such low amounts of pile-up it
is the \gaP\ process that plays a role of the main background thanks to its relatively large cross
section and similar behaviour in the $(m_X, m_{t\bar{t}})$ or proton $(p_{T1}, p_{T2})$ planes - see
Figs.~\ref{proton_pt} and \ref{2Dcut}. However, the $\xi$ distributions of these two processes
differ at low $\xi$ values - see Fig.~\ref{xidist}. A slightly improved significance of 1.2 is
achieved after cutting out the low-$\xi$ part of the spectrum: $0.05 < \xi_{1,2} < 0.15$. For a nearly
same set of cuts (namely no cut on proton $p_T$, $N_{\rm trk} \le 20$ and
$0.05 < \xi_{1,2} < 0.15$) the best significance of 1.1 is reached at the pile-up of
$\langle \mu \rangle =50$. For the HL-LHC luminosity scenario, the maximum significance of 1.2 is
achieved for the $0.2 < p_T^{\rm proton} < 0.7$~GeV and $20 < N_{\rm trk} < 40$ cut configuration, when
the enhanced tracker $|\eta| < 4.0$ coverage is considered. The three cut sets in the
(z-veto region, track $p_T$) space give very similar results.

As we pointed out above, production of the  $t\bar{t}$ system in collision of two Pomerons
suffers from a large uncertainty in the value of $S^2$. To stay on the safe side, we take
conservatively a relatively low value of 3\% for our nominal results above. Doubling this value would
simply double the signal event yield, while leaving all backgrounds unchanged. This would then
lead to significances of 2.4, 2.2 and 1.6 for the three luminosity scenarios, (5, 10), (50, 300) and
(200, 4000), respectively. 

The effective cross-sections after applying individual cuts from
Table~\ref{tab:cuts} and scaling by the rates of fake double tagged events and
by the ToF suppression, are summarized in Table~\ref{tab:cutflow_muall}.

\begin{table}[t]
\begin{tabular}{|c|c|c|c|c|c|c|}
\hline
Process & \gaP($\langle \mu \rangle$=5/50/200) & \PP($\langle \mu \rangle$=5/50/200) & Incl.$t\bar{t}$+PU($\langle \mu \rangle$=5/50/200) \\ \hline
Generated cross section [fb] & 52.0 & 29.4 & 390000.0 \\\hline
$N_{e/\mu} \ge 1~(E_T>25~{\rm GeV}, |\eta|<2.5(4.0))$  & 13.1/13.1/18.3 &6.9/6.9/9.6 & 85138.9/80324.4/129753.0\\ \hline
$N_{\rm jet} \ge 4~(E_T>25~{\rm GeV}, |\eta|<2.5(4.0))$ &4.2/4.6/6.3   &2.0/2.2/4.1 & 33953.4/35150.7/76126.1 \\\hline
$\Delta R(\rm{e/\mu,jet})> 0.2$  &4.2/4.6/6.3 & 2.0/2.2/4.1 &33953.4/35150.7/76116.3 \\\hline
$m_{t\bar{t}}<1000$~GeV, $m_X>400$~GeV &3.9/4.2/6.2 &2.0/2.2/4.1 &28577.3/29406.0/57242.3 \\ \hline
$0.015<\xi_{1,2}<0.15 $           & 2.5/2.5/3.7 &0.7/0.8/1.6 & 147.7/10215.6/54202.7 \\ \hline
ToF suppression & 2.5/2.5/3.7  & 0.7/0.8/1.6  & 8.4/1094.9/21509.0  \\ \hline
\hline
\end{tabular}
\caption{Cut flow for the effective cross sections in femtobarns for the case where the
  semi-exclusive signal process is \gaP\ and the process \PP\ and inclusive $t\bar{t}$ production
  are backgrounds, for three amounts of pile-up, namely $\langle \mu \rangle$ = 5, 50 and 200 (for
  the last, we use the $|\eta| < 4.0$ coverage of the upgraded tracker). 
  The effect of the $\xi$ cut for the inclusive background with pile-up is evaluated as
  a combinatorial background coming from the rate of fake double-tagged events. Suppression of
  pile-up effects from using ToF information is based on $\sigma_{\rm ToF} = 10$~ps and
  Refs.~\cite{Tasevsky:2014cpa,Harland-Lang:2018hmi}.}
\label{tab:cutflow_muall}
\end{table}

A comment about a possible top quark mass measurement using elastic processes is in order. As we stated in
Ref.~\cite{ourfirst}, for a sensible measurement of the top quark mass, one would
need a sufficient amount of signal events and a very low level of background contamination. In the
most promising scenario where the \gaP\ process is separated from all other backgrounds overlaid
with pile-up of $\langle \mu \rangle = 5$ with the significance of 3.4, the signal to background
ratio in terms of event yields from a data sample corresponding to an integrated luminosity of
10~fb$^{-1}$ is 11.1/7.5. Such statistics are still not sufficient to sensibly
complement the top quark mass measurements in inclusive channels. 

\begin{figure}[hbt!]
  \includegraphics[width=0.45\textwidth, height=6cm]{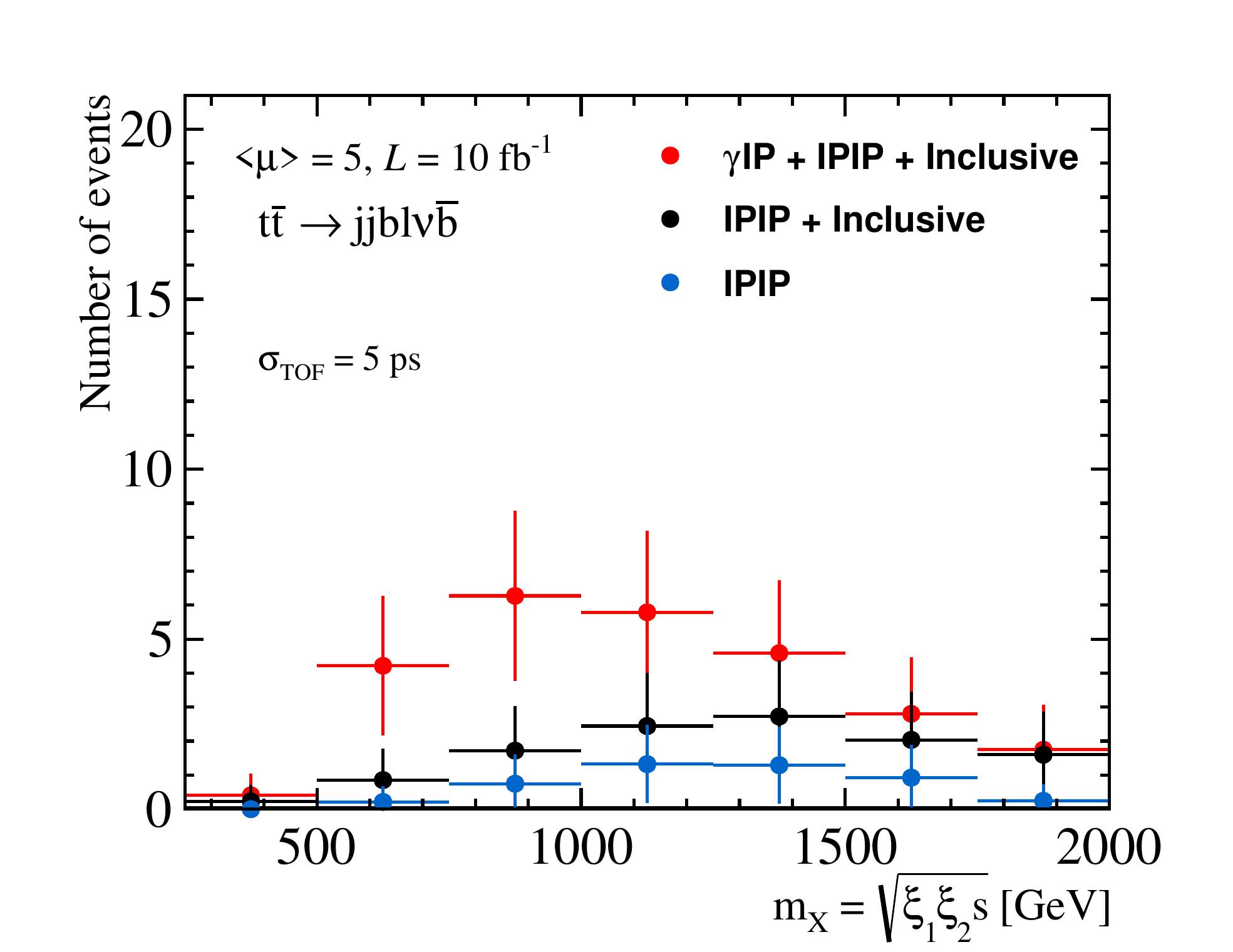}
  \includegraphics[width=0.45\textwidth, height=6cm]{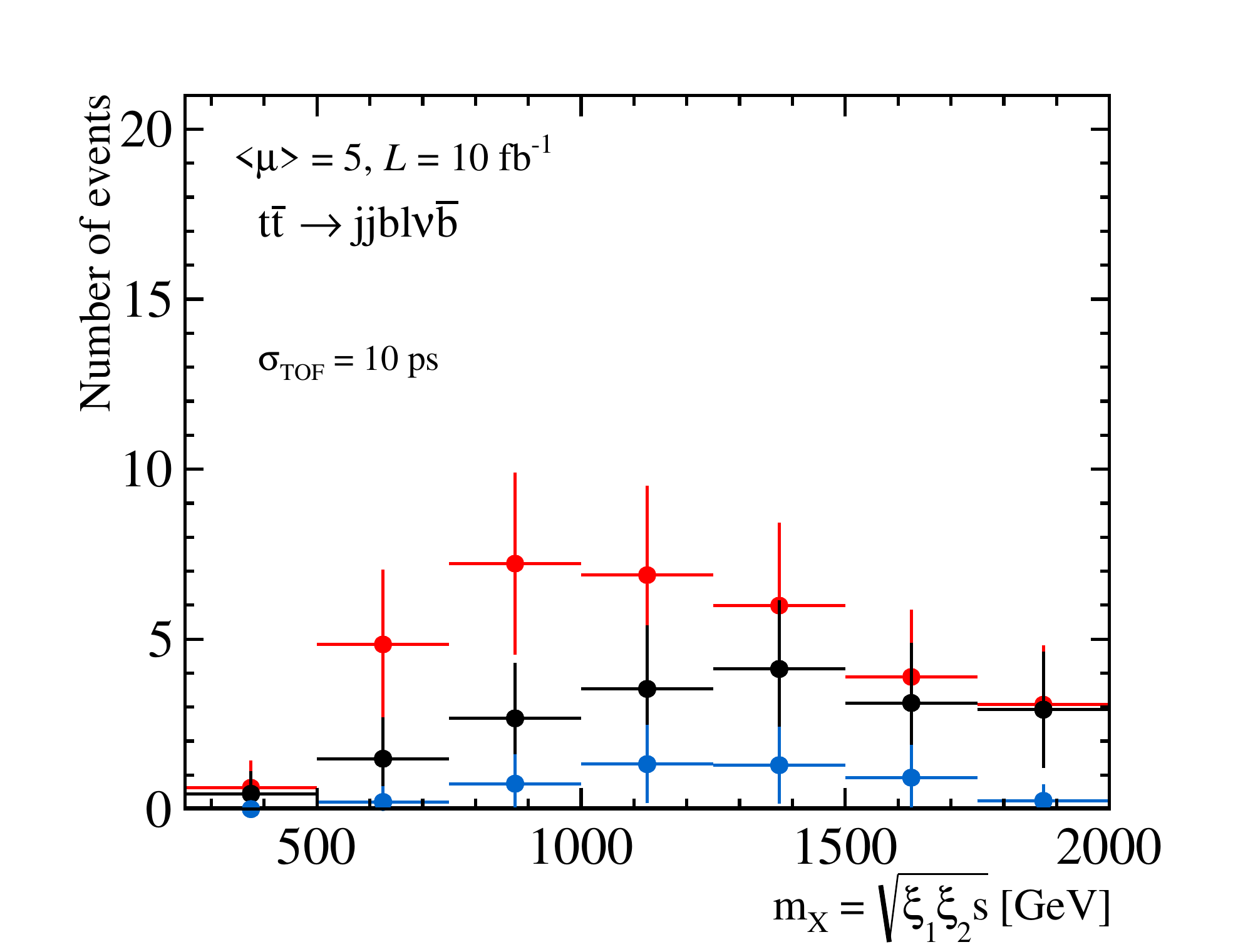}
  \includegraphics[width=0.45\textwidth, height=6cm]{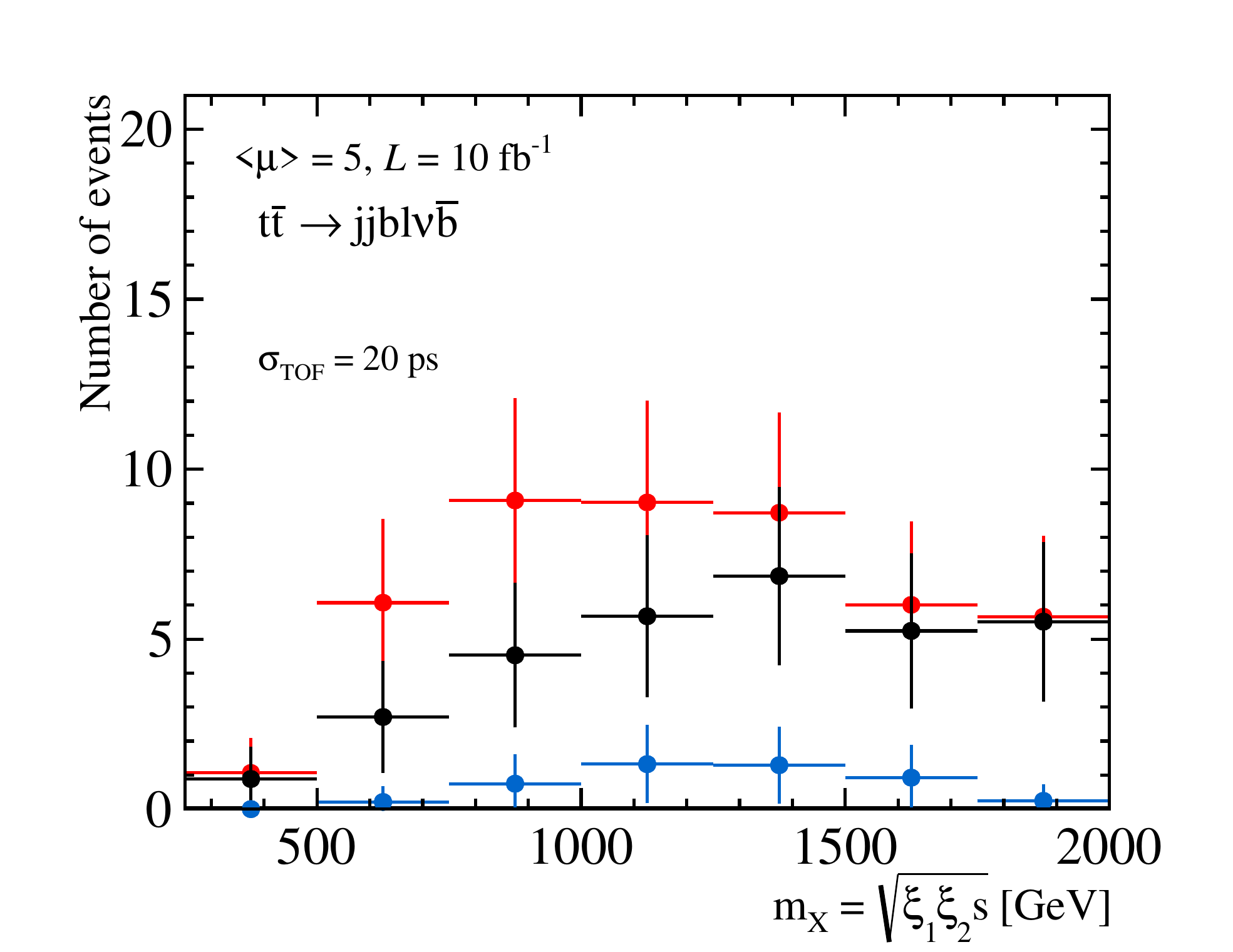}  
  \includegraphics[width=0.45\textwidth, height=6cm]{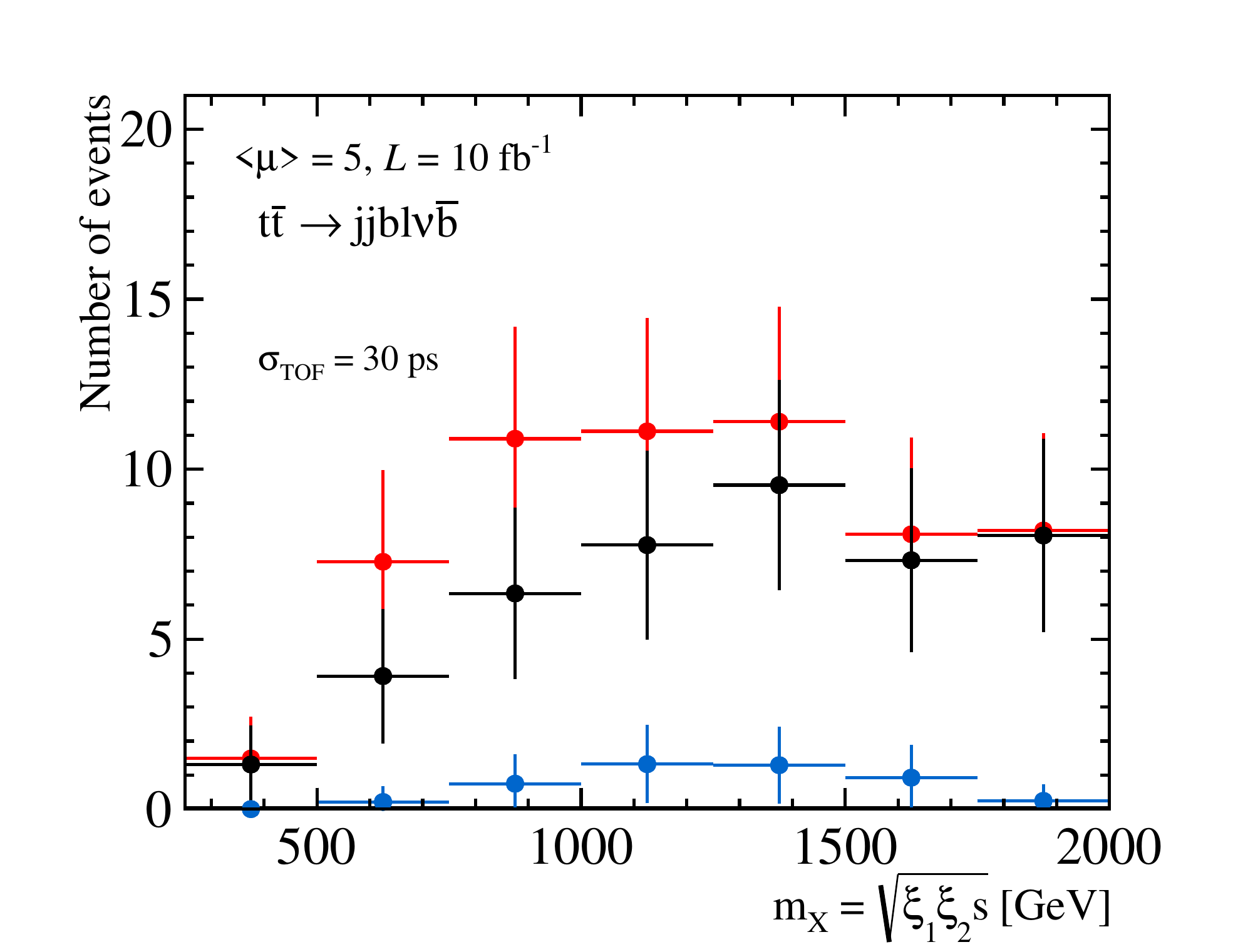}
  \includegraphics[width=0.45\textwidth, height=6cm]{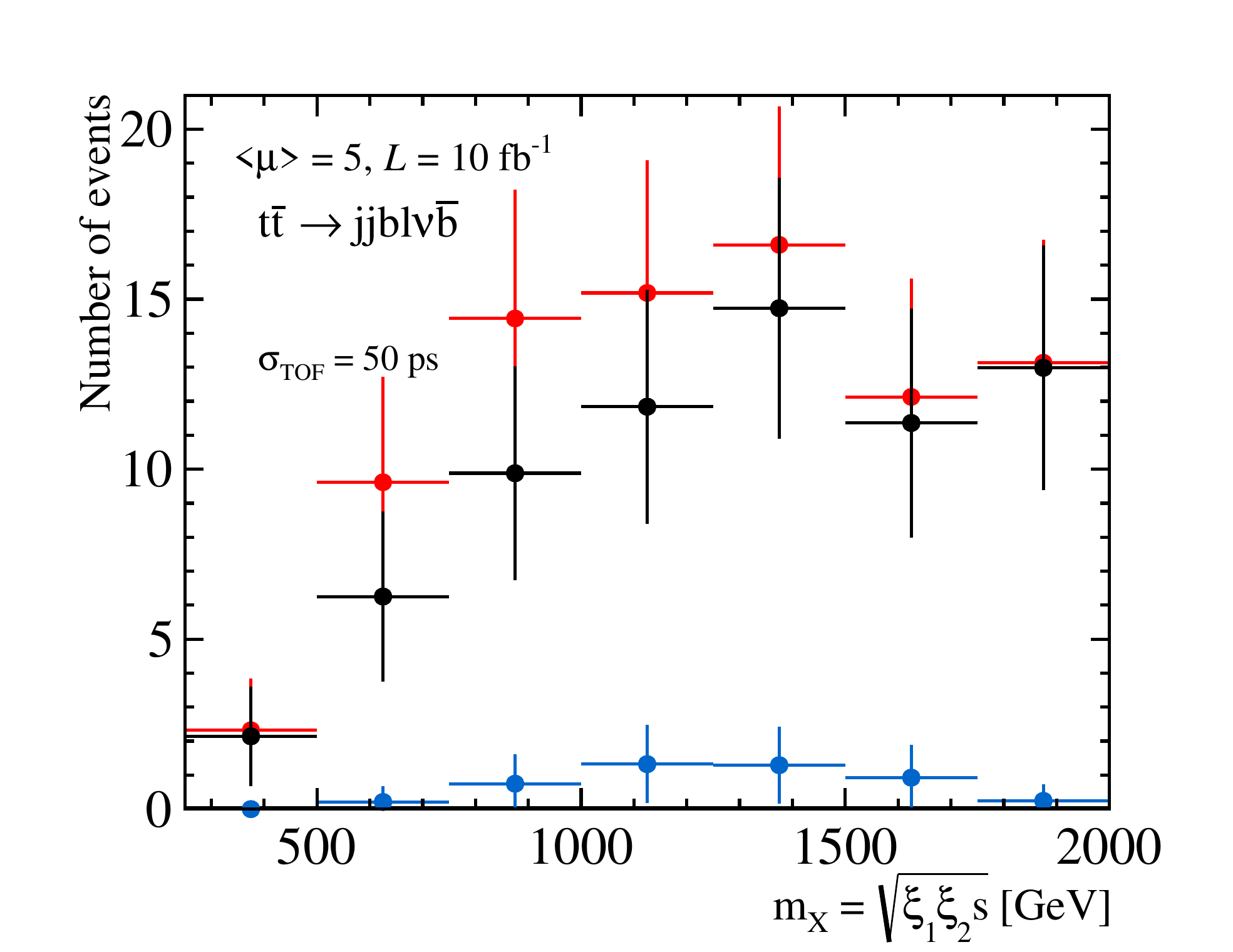}  
  \caption{Separation of \gaP\ as a function of ToF reconstruction, $\sigma_{\rm ToF}$. Distribution
    of missing mass calculated using protons detected in FPDs at generator level
    after applying cuts in Table~\ref{tab:cuts} and the $N_{\rm trk} \leq 20$ cut. For each
    $\sigma_{\rm ToF}$ value, a corresponding suppression
    factor is applied to the inclusive background. Predictions for the two semi-exclusive processes
    \gaP\ and \PP\ are obtained with FPMC, while the inclusive $t\bar{t}$ background was generated with
    MadGraph~5+PYTHIA~8. All are overlaid with pile-up with $\langle \mu \rangle = 5$ interactions
    per event and numbers of events correspond to the integrated luminosity of 10~fb$^{-1}$.}
\label{resolutionstudy_stacked} 
\end{figure}

In what follows, we will study the dependence of our results on the resolution of the ToF detector. This
property of the ToF detector is clearly key for all measurements of elastic processes at LHC.
We remind that AFP reached a time resolution $\sigma_{\rm ToF} =$~20--22~ps in Run~2 (with a rather low,
sub-10\% efficiency~\cite{ToFPUBNote}), whereas CT-PPS time resolution expectations are rather beyond
such values.
While the performance of the ToF detector in general and as functions of pile-up, time resolution and
spatial granularity are studied in detail in Ref.~\cite{ToFperformance}, here we illustrate the effect
of worsening $\sigma_{\rm ToF}$ for one specific process and for one specific (most promising) cut
scenario giving the significance of 3.4 quoted above, i.e. for the process \gaP\ in the luminosity
scenario ($\langle \mu \rangle, \cal L$) = (5, 10) and after applying all cuts in the
Table~\ref{tab:cuts} including the optimal area in the $(m_X, m_{t\bar{t}})$ plane and $N_{\rm trk} \le 20$
cut. While we assume that the number of events for (semi)-exclusive processes does not change with
$\sigma_{\rm ToF}$ at such a small pile-up, the contamination by inclusive process overlaid with
pile--up increases
almost linearly. In numbers this amounts to 11.1 \gaP\ and 1.7 \PP\ events, while the
inclusive background increases from 2.9 events for 5~ps to 5.8 events for 10~ps and 58 events for
50~ps. This would correspond to a gradual deterioration of significance from 4.0 for 5~ps to 3.4, 2.8,
2.3 and 1.4 for 10, 20, 30 and 50~ps, respectively. In the effort to visualize this $\sigma_{\rm ToF}$
dependence, in Fig.~\ref{resolutionstudy_stacked} we plot missing mass spectra $m_X$ for the three
processes and five $\sigma_{\rm ToF}$ values above, but for a slightly different cut scenario, namely
for the whole $(m_X, m_{t\bar{t}})$ area available since this simplifies the projection of the
2D-dependence to the 1D $m_X$-dependence significantly. It should be noted that this simplification
decreases the corresponding significances but not dramatically (e.g. 3.0 for 10~ps rather than 3.4 for
10~ps with the optimal 2D area). 

Finally it is worth reminding that prospects for separation of the sum of \gaP\ and \PP\
processes from all backgrounds remain good, in particular at low pile-up, as emphasized in
Ref.~\cite{ourfirst}. Assuming $\sigma_{\rm ToF} = 10$~ps, maximal significances
reach values of 4.7, 2.7 and 3.0 for the three luminosity scenarios examined in this study,
namely for ($\langle \mu \rangle ,\cal L$) = (5,10), (50, 300) and (200, 4000), respectively, 
where for the last, the enhanced tracker coverage is considered.
The first two are less favourable than those reported in our initial study~\cite{ourfirst} which
is due mainly to two reasons. First, as explained above, we use the more appropriate formula for significance
which is well approximated by the simple ratio $S/\sqrt{B}$ for $S \ll B$ but it gives lower values for
$S\approx B$. Second reason is the use of the most recent version of Delphes. It improves treatment of
low-momentum charged particles which, however, leads to distributions of $N_{\rm trk}$ in signal and inclusive
background process when both overlaid with pile-up closer to each other than we observed in the version
2.5 used in Ref.\cite{ourfirst}. The significance of 3.0 achieved for the HL-LHC conditions promises to keep
the semi-exclusive production of the $t\bar{t}$ system in the portfolio of potentially interesting and
feasible processes for the future of LHC. If the ToF resolution improved to 5~ps at HL-LHC, the statistical
significance would even increase to 3.7. {{This would correspond to a signal event yield of about 16 thousands
and a S/B ratio of the order of one per mille. Extracting such a sample of signal from this relatively
big pile-up contamination is certainly worth the effort but it will require to measure the latter with a
high accuracy (for example by data-driven methods, see e.g.~\cite{AFPll}.)}} 

\section{Summary}
\label{sec:sum}
In recent years, it became clear that the precise study of the top quark production and its decay
at colliders can shed light on several aspects of the SM and provide a way to search for BSM
physics. In general, these studies have been performed considering the top production in inelastic
$pp$ collisions, where the incident protons break up and a large of number of particles is
produced in addition to the top. A cleaner final state is present when the top is produced by the
interaction between color singlet objects emitted by the incident protons. Such elastic top
production can be explored considering the AFP and CT-PPS detectors that are installed
symmetrically around the interaction point at a distance of roughly 210~m from the ATLAS and CMS
detectors. Such a possibility was investigated, for the first time, in Ref.~\cite{ourfirst} and
good prospects for the measurement of this process were obtained. Such results have motivated the
analysis performed here, where we have improved the study considering additional cuts and a more
realistic treatment of the detector response. 

We studied in detail prospects for measuring the $t\bar{t}$ pair produced in the
exclusive (\gaga) and semi-exclusive (\gaP\ and \PP) processes. We analyzed three
luminosity scenarios, going from a low pile-up running ($\langle \mu \rangle = 5$),
through $\langle \mu \rangle = 50$, typical in Run~2 or assumed for the upcoming Run~3,
up to assumed conditions at HL-LHC, $\langle \mu \rangle = 200$. With
the help of Delphes, the main effects of detector acceptance and resolutions
as well as the effect of pile-up background were included in the analysis
procedure. Based on the established selection procedure of the semi-leptonic decay of the 
$t\bar{t}$ pair system, and making use of the exclusive topology of the final state,
the procedure described in Ref.~\cite{ourfirst} has been developed further by searching
for an optimal area in the space of central mass system measured by FPDs
and by central detector and in the space of proton transverse momenta measured on both
sides of FPDs. With this improved selection procedure we investigated potential in
separating the individual (semi)-exclusive processes while considering all remaining ones
as backgrounds.
The inclusive $t\bar{t}$ production overlaid with pile-up remains the most dangerous
background for $\langle \mu \rangle$ of 50 and 200. 
Although compared to results of Ref.~\cite{ourfirst} significances for individual processes
improved, still they do not exceed values of 0.1 in the case of the \gaga\ process whose
cross section is predicted to be extremely low, and values of 2.0 in the case of the \PP\
process whose cross section is about a half of that for the \gaP\ process but whose
characteristics in key observables are strikingly similar to those for \gaP. The best significance
of 3.4 was obtained for the \gaP\ process for the lowest pile-up scenario giving thus
promising prospects to study this process, which is theoretically well under control, for
the first time at LHC and in much detail. Outlooks are also favorable for extraction of the sum
of the \gaP\ and \PP\ processes. While significance reaches almost 5.0 for low pile-up
contaminations, its observation seems to be possible at HL-LHC conditions. 

Our results indicate that the discovery of the elastic top pair production in $pp$ collisions at the LHC is feasible and that it is possible to probe, for the first time, the diffractive photoproduction of top quark pairs. Such a perspective strongly motivates the search for new physics beyond the Standard Model in this process, which we plan to discuss in a forthcoming publication. We are convinced that studying
the semi-exclusive production of such a complex system as the $t\bar{t}$ pair can be considered a solid
part of the physics programme of forward proton detectors at LHC and HL-LHC.


\begin{acknowledgments}
  The authors acknowledge a useful discussion with Michelangelo Mangano during the LHC Top WG
  meeting and thank Michelle Selvaggi for the help with the Delphes card used by the CMS
  Collaboration at HL-LHC. This work was partially financed by the Brazilian funding agencies
  CNPq (process number 164609/2020-2), FAPERGS and INCT-FNA (process number 464898/2014-5).
  MT is supported by MEYS of the Czech Republic within project LTT17018.
\end{acknowledgments}

\end{document}